\documentclass[final,3p,times,twocolumn]{elsarticle}

\usepackage{graphicx}
\usepackage{amsmath}
\usepackage{color}

\journal{High Energy Density Physics}

\begin{document}

\begin{frontmatter}





\title{Dynamics of volumetrically heated matter\\passing through the liquid--vapor metastable states}
\author[address1]{Steffen Faik\corref{cor1}}
\address[address1]{Institut f\"ur Theoretische Physik, Goethe-Universit\"at Frankfurt am Main, Max--von--Laue--Str. 1, 60438 Frankfurt am Main, Germany}
\cortext[cor1]{Corresponding author. Address: Institut f\"ur Theoretische Physik, Goethe-Universit\"at Frankfurt am Main, Max--von--Laue--Str. 1, 60438 Frankfurt am Main, Germany. Tel.: +49 (0)69 798 47846.}
\ead{faik@th.physik.uni-frankfurt.de}
\ead[url]{http://th.physik.uni-frankfurt.de/$\sim$faik/}
\author[address2,address3]{Mikhail M. Basko}
\address[address2]{ExtreMe Matter Institute EMMI, GSI Helmholtzzentrum f\"ur Schwerionenforschung GmbH, Planckstr. 1, 64291 Darmstadt, Germany}
\address[address3]{Institute for Theoretical and Experimental Physics (ITEP), Bolshaya Cheremushkinskaya 25, 117218 Moscow, Russia}
\author[address1,address2]{Anna Tauschwitz}
\author[address2,address4,address5]{Igor Iosilevskiy}
\address[address4]{Joint Institute for High Temperature of RAS, Izhorskaya str. 13 k 2, 125412 Moscow, Russia}
\address[address5]{Moscow Institute of Physics and Technology, Aerophysics and Space Research Department, Institute lane 9, 141700 Dolgoprudny, Moscow region, Russia}
\author[address1,address2]{Joachim A. Maruhn}

\begin{abstract}
Remaining within the pure hydrodynamic approach, we formulate a self-consistent model for simulating the dynamic behavior of matter passing through metastable states in the two-phase liquid--vapor region of the phase diagram. The model is based on the local criterion of explosive boiling, derived by applying the theory of homogeneous bubble nucleation in superheated liquids. Practical application of the proposed model is illustrated with hydrodynamic simulations of a volumetrically uniformly heated planar layer of fused silica SiO${}_2$. Implications for experimentally measurable quantities are briefly discussed. A newly developed equation of state, based on the well known QEOS model and capable of handling homogeneous mixtures of elements, was used in the numerical simulations.
\end{abstract}


%% \pacs{47.40.-x, 64.10.+h, 64.60.My, 64.70.F-}

\end{frontmatter}

\section{\label{s:intro} Introduction}

Most experiments with warm dense matter are intimately coupled to an adequate analysis of the hydrodynamic motion of the studied material, and the solution of the hydrodynamic equations requires knowledge of the thermodynamic equation of state (EOS). In many cases --- unless the external energy deposition goes rapidly well beyond the latent heat of vaporization --- the heated (or cooled) material passes through the two-phase region of the phase diagram, where the liquid and the vapor phases can coexist (for simplicity, we do not consider the solid phase here). Part of this region (between the binodal and the spinodal) corresponds to metastable states, where the thermodynamically consistent EOS, based on the reversibility considerations, is essentially double-valued and poses the following dilemma for any hydrodynamic simulation: one can either follow the metastable branch (the MS-EOS) of the superheated liquid (supercooled vapor), or use the fully equilibrium EOS (the EQ-EOS) obtained by the Maxwell construction across the van der Waals loops. In this paper, by using a simple example of a uniformly heated planar foil, we show how this dilemma can be resolved within the framework of hydrodynamics, without invoking much more complex and less clearly defined equations for the kinetics of phase transition.

Our basic proposition is to follow the evolution of the heated liquid into the two-phase region by applying the MS-EOS until the criterion of explosive boiling is fulfilled. Once the boiling point is reached, we assume that the considered fluid element undergoes a practically instantaneous and irreversible transition to the EQ-EOS. The criterion of explosive boiling (\ref{bc:b_crit=}), which plays the central role in our model, is derived by applying the theory of homogeneous bubble nucleation in superheated liquids \cite{Skr74}.

The proposed method is applied to a thin planar layer of fused silica (SiO${}_2$) that is volumetrically uniformly heated  at a rate of $10^{11}$~J~g${}^{-1}$~s${}^{-1}$ on a time scale of 100~ns. Earlier this configuration was proposed \cite{TaEf.09} to be used in warm-dense-matter experiments under irradiation by intense beams of heavy ions at GSI\footnote{GSI Helmholtzzentrum f{\"u}r Schwerionenforschung GmbH, Darmstadt, Germany, http://www.gsi.de. GSI operates a unique large-scale accelerator for heavy ions.}. Initially, in its solid--liquid state, the SiO${}_2$ foil is so thin that its thermal expansion proceeds in a quasi-static manner, so that the pressure across the foil rapidly relaxes to its boundary value. For illustrative purposes, we assumed in simulations a boundary pressure of $p_0 =500$~bar. But even if in real experiments much lower boundary pressures can be anticipated, our basic conclusions should remain valid because this value of $p_0$ is already quite small compared to the critical pressure.

Although in this study only a very simple and perfectly uniform target configuration has been analyzed, the proposed method of passing through the two-phase region can be applied to a much broader spectrum of experimental conditions. How well suited it turns out to be for modeling these conditions, remains to be clarified by future work.

This paper is organized as follows. In section~\ref{s:qs} the quasi-static regime of expansion of a volumetrically heated planar foil is analyzed in detail. In section~\ref{s:os} possible role of quasi-elastic oscillations is discussed. The criterion for explosive boiling is derived and discussed in detail in section~\ref{s:bc}. The results of hydrodynamic simulations, which illustrate practical application of the proposed method for treatment of the two-phase region, are presented in section~\ref{s:res}. All the hydrodynamic simulations have been performed by using a newly developed EOS package FEOS, whose brief description is given in \ref{s:MPQeos}.

\section{\label{s:qs} Quasi-static thermal expansion of a planar foil}

Suppose that an initially uniform planar foil at a density $\rho_0$  is subject to an external volumetric heating with a prescribed energy deposition rate $q$ (energy per unit mass per unit time). We assume for simplicity that $q$ may depend on time $t$ (starting from $t=0$) but not on the spatial coordinate $x$, measured perpendicular to the foil surface. The ambient (boundary) pressure is fixed at some positive value $p_0>0$, which is assumed to be small compared to the critical pressure of the liquid--vapor phase transition. In the course of heating the foil undergoes expansion, and its thickness $l =l(t)$  increases in time starting from the initial value $l_0= l(0)$.

Clearly, if the foil is heated slowly by comparison with the sonic time
\begin{equation}\label{qs:t_s=}
  t_s = \frac{l}{c_s},
\end{equation}
where $c_s$ is the speed of sound, it expands quasi-isobarically in a quasi-static manner. The solution of the hydrodynamic equations, describing the quasi-static regime of expansion, is readily obtained by successive approximations.

In the zeroth approximation pressure $p$ is constant in space and time
\begin{equation}\label{qs:p=}
  p(t,x) =p(\rho,T) = p_0,
\end{equation}
while the heating rate $q$ determines temporal variation of the specific enthalpy $h$ according to the basic thermodynamic law
\begin{equation}\label{qs:dh/dt=}
  \frac{dh}{dt} =q,
\end{equation}
which yields
\begin{equation}\label{qs:h=}
  h(t,x) =h(t) = h(\rho,T) =h_0 + \int\limits_0^{t} q(\tau)\, d\tau.
\end{equation}
From the two equations (\ref{qs:p=}) and (\ref{qs:h=}) one knows the zero-order values of all thermodynamic variables at any given time $t>0$; their spatial distribution remains perfectly uniform. The fluid velocity $u$ in the zeroth approximation is zero.

Substituting the zero-order fluid density $\rho =\rho(t)$ into the mass conservation law
\begin{equation}\label{qs:hyd1=}
  \frac{\partial\rho}{\partial t} +
  \frac{\partial(\rho u)}{\partial x}  =0,
\end{equation}
we obtain the first-order solution for the fluid velocity
\begin{equation}\label{qs:u_q=}
  u(t,x)=u_q(t,x) =ax,
\end{equation}
where
\begin{equation}\label{qs:a(t)=}
  a =a(t) =-\frac{d\ln\rho}{dt} =\frac{d\ln v}{dt}=\frac{d\ln l}{dt};
\end{equation}
here and below $v = \rho^{-1}$ is the specific volume. In Eq.~(\ref{qs:u_q=}) we assumed that the value $x=0$ corresponds to the motionless center of mass, i.e.\ to the middle of the foil.

By virtue of Eq.~(\ref{qs:dh/dt=}), the expansion rate $a$ can be cast in a form
\begin{equation}\label{qs:a=}
  a = -q \left(\frac{\partial h}{\partial\ln\rho}\right)^{-1}_p =\Gamma\,\frac{ q}{c_s^2},
\end{equation}
where the dimensionless coefficient $\Gamma$ --- which is just a thermophysical parameter of the studied material --- is defined as
\begin{equation}\label{qs:Gam=}
  \Gamma = -c_s^2 \left(\frac{\partial h}{\partial\ln\rho}\right)^{-1}_p.
\end{equation}
One readily verifies that for a perfect gas $\Gamma =\gamma-1$, where $\gamma$ is the adiabat index. For a two-term EOS of the form
\begin{equation}\label{qs:2term_EOS=}
  \begin{array}{l}
  \epsilon(\rho,T) =\epsilon_c(\rho) +c_V T, \\
  p(\rho,T) =\displaystyle \rho^2\frac{d\epsilon_c}{d\rho}
  +\Gamma_G \,\rho c_V T,
  \end{array}
\end{equation}
often applied to describe solids and liquids at high energy densities \cite{ZeRa02}, our $\Gamma$ coefficient coincides with the Gr\"uneisen parameter $\Gamma_G$ --- provided that the latter is assumed to be constant; in Eq.~(\ref{qs:2term_EOS=}) $\epsilon=h-p/\rho$ is the mass-specific internal energy, $c_V$ is a constant heat capacity at fixed volume.

The physical meaning of the expansion rate $a$ is straightforward: $a^{-1}$ is the time scale over which a quasi-isobaric heating results in a significant change of the density $\rho$. Then, the obvious necessary condition for applicability of the quasi-static solution (\ref{qs:p=}), (\ref{qs:h=}) and (\ref{qs:u_q=}) is
\begin{equation}\label{qs:at_s<=}
  at_s \ll 1.
\end{equation}
In other words, the dimensionless product $at_s$ is the small parameter which defines the hierarchy of successive approximations in our solution of the hydrodynamic equations. From Eq.~(\ref{qs:u_q=}) we find the quasi-static expansion velocity $u_{ql}$ of the foil boundary at $x=l/2$
\begin{equation}\label{qs:u_ql=}
  u_{ql} = \frac{1}{2}\, al = \frac{c_s}{2} \, (at_s),
\end{equation}
which, as might be expected, is a first-order small quantity compared to the sound speed $c_s$.

Inequality (\ref{qs:at_s<=}) is certainly the necessary but not always a sufficient condition for practical application of the quasi-static solution. Having substituted the first-order velocity (\ref{qs:u_q=}) into the hydrodynamic equation of motion
\begin{equation}\label{qs:hyd2=}
  \frac{\partial u}{\partial t} +u\frac{\partial u}{\partial x} +\rho^{-1} \frac{\partial p}{\partial x} =0,
\end{equation}
one finds that this solution is generally characterized by small spatial pressure non-uniformities of the order
\begin{equation}\label{qs:del_p=}
  \delta p \simeq \rho u_{ql}^2 \simeq \rho c_s^2 \, (at_s)^2.
\end{equation}
Because the mass-specific heating rate $q$ is assumed to be perfectly uniform, the same refers to the mass-specific entropy $s$ at all times, and the resulting spatial non-uniformity of the density can be evaluated as
\begin{equation}\label{qs:del_rho=}
  \delta\rho= \delta p
  \left(\frac{\partial p}{\partial\rho}\right)^{-1}_s =
  \frac{\delta p}{c_s^2} \simeq \rho\, (at_s)^2.
\end{equation}
Thus, the quasi-static solution (\ref{qs:p=}), (\ref{qs:h=}), (\ref{qs:u_q=}) maintains spatial uniformity down to the second order in $at_s$.

Now, if a particular experiment aims at measuring the density difference
\begin{equation}\label{qs:Del_rho=}
  \Delta\rho = \rho(t+ \Delta t)-\rho(t) \simeq \rho \,a\Delta t
\end{equation}
between two quasi-uniform states, separated by a time interval $\Delta t$, the measured difference must be large in comparison with the inherent spatial non-uniformity, i.e.\ $\Delta\rho \gg \delta\rho$, which implies
\begin{equation}\label{qs:Dt>>=}
 \Delta t \gg t_s\, (at_s).
\end{equation}
From practical point of view condition (\ref{qs:Dt>>=}) is not particularly restrictive because it allows application of the quasi-static solution even on time scales $\Delta t \lesssim t_s$ shorter than the sonic time --- provided, of course, that no significant quasi-elastic oscillations have been excited (see next section).

In practice, the basic condition (\ref{qs:at_s<=}) for quasi-static approximation can be satisfied with relatively high heating rates $q \simeq 10^{11}$~J~g${}^{-1}$~s${}^{-1}$, typical for irradiation by heavy ion beams at GSI, if a sufficiently thin foil is used. In a typical solid the above value of $q$ implies an expansion rate of $a\simeq 10^7$~s${}^{-1}$, and the product $at_s$ falls in the range $at_s \lesssim 0.03$ for foils with $l \lesssim 10$~$\mu$m.

If a quasi-statically heated foil remains in a liquid state and preserves a sharp surface, its expansion velocity $u_{ql}$ (which even under quasi-static conditions may amount to 10--100~m~s${}^{-1}$) is an easily measurable quantity, which provides direct information about the density $\rho$ at any given time $t\gg t_s$. Once the energy deposition $\int q\, dt$ is known and the surface temperature $T$ has been measured, one obtains the thermal, $\alpha_p =-\rho^{-1}\left(\partial\rho/\partial T\right)_p$, and the caloric, $\alpha_p^{\ast} =-\rho^{-1}\left(\partial\rho/\partial h\right)_p$ expansion coefficients together with the heat capacity $c_p= \left(\partial h/\partial T\right)_p$ at constant pressure.

A more advanced experimental setup, consisting of several foils, is described in Ref.~\cite{TaEf.09}. The general goal is to employ volumetrically heated samples for EOS measurements near the boiling curve \cite{IoGr05}. In particular, the subject of non-congruent phase transitions can be addressed. Non-congruence means coexistence of two (or more) phases with different chemical composition (stoichiometry). One of the reasons why SiO${}_2$ has been proposed as a first candidate for such experiments is that it is expected to manifest non-congruent evaporation \cite{IoHy.01,RoIo.04}.

\section{\label{s:os} Quasi-elastic oscillations and initialization of the hydro simulation}

An obvious problem with the quasi-static solution (\ref{qs:p=}), (\ref{qs:h=}), (\ref{qs:u_q=}) is that one cannot in general satisfy the condition of zero initial velocity $u(0,x)=0$. If, for example, the maximum heating rate is turned on at once at $t=0$,
\begin{equation}\label{os:q=}
  q(t) = \left\{ \begin{array}{ll} 0, & t<0, \\
  q_m, & t \geq 0, \end{array} \right.
\end{equation}
Eq.~(\ref{qs:u_q=}) does not allow to start from zero initial velocity --- even when $q_m$ is relatively small and condition (\ref{qs:at_s<=}) is satisfied. The reason lies in the fact that Eq.~(\ref{qs:u_q=}) is not the general but a particular first-order solution of the hydrodynamic equations for small-amplitude motions with $u/c_s \sim at_s <<1$. To obtain a general solution, one has to take into account quasi-elastic oscillations, represented by an infinite series of eigenmodes for standing acoustic waves; see \ref{s:Ac}.

\begin{figure}
\centering
\includegraphics*[width=\columnwidth]{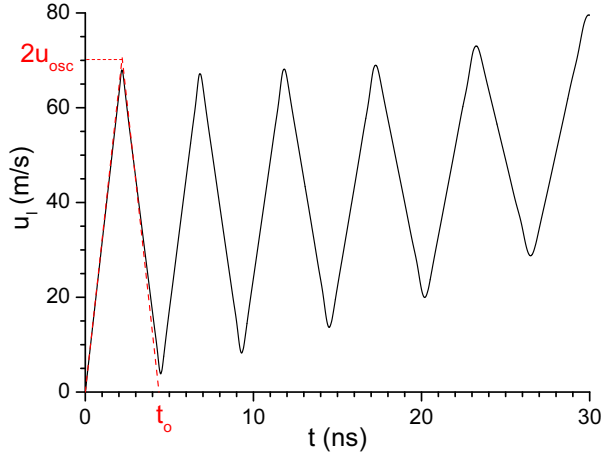}
\caption{\label{f:os} Temporal variation of the surface velocity $u_l$ of a 10~$\mu$m thick SiO${}_2$ planar foil by a sudden turnon of the volumetric heating at a rate $q= q_m= 10^{11}$~J~g${}^{-1}$~s${}^{-1}$. The amplitude of quasi-elastic oscillation is $u_{osc}= 3528$~cm/s.}
\end{figure}

Fig.~\ref{f:os} shows the time variation of the surface velocity $u_l$ by a sudden turnon of the volumetric heating as calculated numerically with the CAVEAT code. Initially, the 10~$\mu$m thick SiO${}_2$ foil was at rest. One clearly sees that periodic oscillations with an amplitude $u_{osc}$ are superimposed on a slow ``secular'' expansion of the foil. It can be rigorously proven (see \ref{s:Ac}) that at $t \ll a^{-1}$ such oscillations have a period $t_o =2l_0/c_s$ and an amplitude $u_{osc}= u_{ql} =\frac{1}{2}al_0$.

Clearly, for some experimental tasks quasi-elastic oscillations might be useful (to infer, for example, the sound speed $c_s$ from the measured period $t_o$ of oscillations), for others --- highly undesirable because they introduce much larger (first-order in $at_s$) spatial variations of pressure and density than are inherent in the quasi-static solution. In particular, for sufficiently low boundary pressures $p_0 < \rho_0 c_s^2 \,(at_s)$, the interior fluid pressure becomes negative in the contraction phase.

In this work we deal with a situation where quasi-elastic oscillations are unwanted because they only introduce an unnecessary complication into our illustrative simulations discussed below. Experimentally, oscillations can be suppressed by raising the heating rate $q$ smoothly from zero to its maximum level over a time scale of $\Delta t \simeq a^{-1}$. In theoretical analysis they can be eliminated by simply assigning the linear initial velocity profile (\ref{qs:u_q=}), predicted by the quasi-static solution, as it has been done in the present work.

\section{\label{s:bc} Criterion for explosive boiling}

\subsection{General considerations}

\begin{figure}
\centering
\includegraphics*[width=\columnwidth]{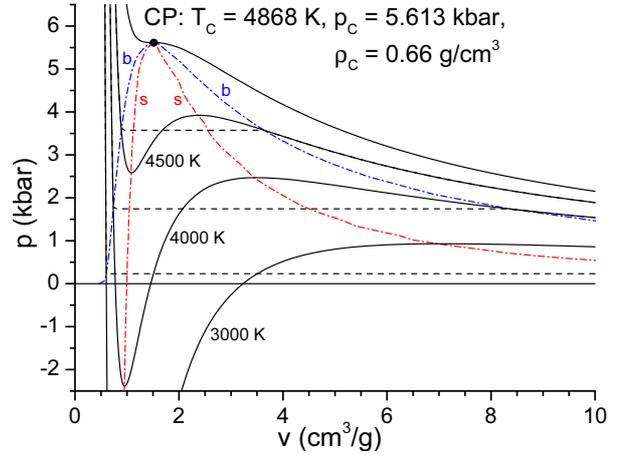}
\caption{\label{f:SiO2_EOS} Equation of state of SiO$_2$ according to the FEOS model on the volume--pressure phase plane. Shown are the isotherms with (the EQ-EOS) and without (the MS-EOS) the Maxwell construction; the binodal and the spinodal touch each other at the critical point CP.}
\end{figure}

Here we explain our principal idea on how to treat the hydrodynamics of metastable states on the example of a thin planar foil of SiO${}_2$. The realistic EOS of SiO${}_2$, used in the hydrodynamic simulations, is shown in Fig.~\ref{f:SiO2_EOS} in standard representation on the $(v,p)$ plane and described in detail in~\ref{s:MPQeos}. It has all the characteristic features of the van der Waals EOS: the boiling and the condensation curves make up the binodal; the region of absolute thermodynamic instability is delimited by the spinodal, which lies below the binodal and touches it at the critical point.

In our example the initially uniform SiO${}_2$ foil enters the metastable region from the liquid side in a quasi-static manner due to a uniform volumetric heating. Application of the quasi-static approximation simplifies the analysis because, at least initially, the entire foil is represented by a single point on the thermodynamic diagrams. It should be noted however that the proposed algorithm for traversing the double-valued EOS region has general applicability for any dynamic non-uniform fluid, entering the metastable region from either the liquid or the vapor sides.

\begin{figure}
\centering
\includegraphics*[width=0.9\columnwidth]{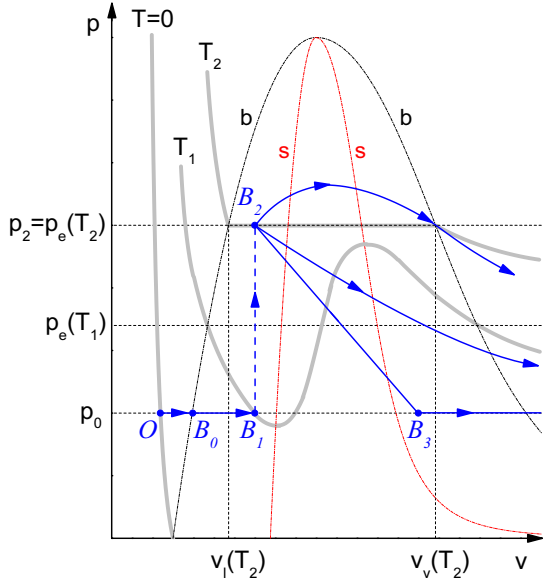}
\caption{\label{f:MS->EQ} Schematic representation of the evolution of the thermodynamic state of heated matter on the $(v,p)$ phase plane.  $B_1\rightarrow B_2$: transition to the phase equilibrium.}
\end{figure}

Fig.~\ref{f:MS->EQ} gives a schematic view of the trajectory of the heated foil material on the $(v,p)$ plane. We can assume that it starts from point $O$ on the cold curve $T=0$ from the minimum possible  specific volume at a given pressure $p=p_0$. Having crossed the liquid binodal at point $B_{\textrm{0}}$, it continues into the metastable region along the liquid MS-EOS branch because at point $B_{\textrm{0}}$ the boiling time is formally still infinitely large.

As the $(v,p)$ state of the heated foil penetrates deeper and deeper into the metastable region of the superheated liquid, the thermodynamic barrier for transition into a fully equilibrium state, prescribed by the EQ-EOS, is lowered and, as a consequence, the time scale for transition into a new state very rapidly decreases by many orders of magnitude. As a result, at a certain point $B_1$ one observes very rapid (explosive) boiling of a superheated liquid from inside, irrespective of the boundary conditions, i.e.\ even in an infinitely large volume of a perfectly uniform and pure liquid \cite{Skr74}.

Keeping this qualitative picture in mind, our basic proposition is as follows: the thermodynamic evolution of every fluid element is followed along the MS-EOS until the criterion for explosive boiling (rapid condensation) is fulfilled at a certain point $B_1$ between the binodal and the spinodal; at point $B_1$ one makes an instantaneous transition from the MS-EOS to the EQ-EOS (to the point $B_2$ in Fig.~\ref{f:MS->EQ}) at constant values of the fluid density $\rho =v^{-1}$ and the mass-specific internal energy $\epsilon$. It is presumed that the criterion itself is local, i.e.\ does not depend on the boundary conditions and the thermodynamic state of adjacent fluid elements.

Our rationale for keeping precisely the density and the specific internal energy constant during the $\mbox{MS} \rightarrow \mbox{EQ}$ transition is based on the fact that the time scales for the $\rho$ and $\epsilon$ change in the considered class of phenomena are governed by the hydrodynamics and the rate of external heating (cooling). Both these time scales are usually much longer than that of the explosive boiling (if not at the transition point itself, then immediately beyond it at a slightly higher temperature; see Fig.~\ref{f:tau_b} below). For comparison, near the spinodal the rate of  transition to the equilibrium state can roughly be estimated as a few times the mean intermolecular distance divided by the mean thermal velocity of molecules, i.e.\ around $10^{12}$~s${}^{-1}$ --- which is practically instantaneous by any hydrodynamic measure.

Fixed values of $\rho$ and $\epsilon$ imply that the $B_1 \rightarrow B_2$ transition will be accompanied by an instantaneous jump in the values of pressure, temperature and specific entropy. According to the general law of increase of entropy \cite{LL-sp96}, the final state $B_2$ must always have a higher specific entropy than the initial state $B_1$  because any fluid element at fixed $\rho$ and $\epsilon$ can be considered as a closed thermodynamic system undergoing a transition from a metastable state to a more probable fully equilibrium state. Less obvious is the direction of the temperature change. We expect that normally the temperature should also jump to a higher value $T_2>T_1$ (as it does in the FEOS model used in this work), although we have managed to prove this fact rigorously only for the analytical van der Waals EOS itself (see \ref{s:T2}).

\subsection{Mathematical formulation}

To express the  criterion for explosive boiling in mathematical terms, we apply the theory of homogeneous bubble nucleation in superheated liquids \cite{Skr74,BlKa75}. According to this theory, thermodynamic fluctuations in a superheated liquid lead to a spontaneous creation of vapor bubbles with different radii $r$. Analysis of the mechanical and chemical equilibrium of these bubbles reveals that there is a critical value $r_c$ of the bubble radius, defined by the condition
\begin{equation}\label{bc:r_c=}
  p_e = p+\frac{2\sigma}{r_c},
\end{equation}
where $\sigma$ is the surface tension, and $p_e =p_e(T)$ is the saturated vapor pressure at temperature $T$ (see Fig.~\ref{f:MS->EQ}); here and below the thermodynamic parameters $\rho$, $T$, $p$ and $\epsilon$ without indices refer to the superheated liquid obeying the MS-EOS. Small bubbles with $r< r_c$ are dominated by surface tension and rapidly collapse, giving no contribution to the process of boiling. Bubbles with $r\geq r_c$ grow and add to the volume fraction occupied by the vapor. Hence, the boiling dynamics is governed by the rate $J$ [cm${}^{-3}$~s${}^{-1}$] of spontaneous creation of vapor bubbles with the critical radius $r_c$.

By its physical meaning, the nucleation rate $J$ defines the time scale of boiling of a superheated liquid: if we fix a certain fluid volume $V$ and wait for time $t$, then on average $JVt$ critical vapor bubbles will be spontaneously created in this volume. Hence, if one knows that in full equilibrium the vapor phase occupies a volume fraction $\xi_v$ and that, in addition, the subsequent growth of newly created critical bubbles can be ignored, one readily calculates that under stationary conditions the transition from the metastable to the absolutely stable state occurs after a waiting time
\begin{equation}\label{bc:t_b=}
    t_b =\xi_v (JV_c)^{-1},
\end{equation}
where $V_c=4\pi r_c^3/3$ is the volume of a critical bubble.
Below we show that in the dynamic situations analyzed in the present work, where the evolution of the thermodynamic state of every fluid element is governed by a combination of hydrodynamics and external heating (cooling), the growth of supercritical bubbles can indeed be neglected, and that the effective waiting time is given by
\begin{equation}\label{bc:t_b==}
    t_b= \left[\frac{d}{dt} \left(-\frac{W_c}{T} \right)\right]^{-1}.
\end{equation}

The key quantity for evaluating the nucleation rate $J$ is the  minimum work
\begin{equation}\label{bc:W_c=}
  W_c= \frac{4\pi}{3} \, \sigma  r_c^2 =
  \frac{16\pi \sigma^3}{3(p_e-p)^2}
\end{equation}
needed to create a critical bubble.  Eqs. (\ref{bc:r_c=}) and (\ref{bc:W_c=}) ignore the small Poynting correction to $p_e-p$, discussed in Ref.~\cite{BlKa75}. By combining the arguments of physical kinetics and thermodynamics, different versions of the basic formula for $J$ have been derived and investigated by different authors \cite{Skr74}, of which probably the simplest, but quite adequate for our case, will be \cite{BlKa75}
\begin{equation}\label{bc:J=}
  J = N \left(\frac{3\sigma}{\pi m } \right)^{1/2} \exp\left(-\frac{W_c}{T}\right),
\end{equation}
where $N=\rho/m$ is the number of liquid molecules per unit volume, and $m$ is the molecular mass.

We determine the moment $t$ of the $\mbox{MS} \rightarrow \mbox{EQ}$ transition from the condition that, when a fluid element jumps from a metastable liquid state $(\rho,T)=(\rho,T_1)$ to an equilibrium two-phase state $(\rho,T_2)$ (see Fig.~\ref{f:MS->EQ}), the total fractional volume of all vapor bubbles with supercritical radii, accumulated in the fluid by that moment, reaches its equilibrium value
\begin{equation}\label{bc:xi_v=}
 0\leq \xi_v =  \xi_v(\rho,T) = \frac{\rho_l(T_2)-\rho}{\rho_l(T_2)-\rho_v(T_2)} \leq 1,
\end{equation}
prescribed by the EQ-EOS for the  final state $(\rho,T_2)$; here $\rho_l(T)$ and $\rho_v(T)$ are, respectively, the liquid and the vapor densities along the corresponding branches of the binodal (see Fig.~\ref{f:MS->EQ}). The post-transition temperature $T_2 =T_2(\rho,T)$ is implicitly defined by the equation
\begin{equation}\label{bc:T_2=}
  \epsilon(\rho,T) =\epsilon_{EQ}(\rho,T_2).
\end{equation}
In other words, for a given EOS of the van der Waals type, $T_2(\rho,T)$ is a known function of the density $\rho$ and temperature $T$ in the metastable state. Normally one has $T_2(\rho,T)> T$; for the van der Waals EOS itself the mathematical proof of this fact is given in \ref{s:T2}.

After we take into account that a bubble, created at a time $t'< t$ with a critical volume $V_c(t') =4\pi r_c^3(t')/3$, by the time $t$ will occupy the volume
\begin{equation}\label{bc:V=}
  V(t) = V_c(t') +\int\limits_{t'}^t \dot{V}_c(t',t'')\, dt'',
\end{equation}
we obtain the following mathematical form of the above condition for the fractional vapor volume
\begin{equation}\label{bc:b_crit0=}
  \int\limits_0^t J(t') \left[V_c(t') + \int\limits_{t'}^t \dot{V}_c(t',t'')\, dt''\right] \, dt' =\xi_v.
\end{equation}
 In Eqs. (\ref{bc:xi_v=}) and (\ref{bc:b_crit0=}) $\dot{V}_c(t',t'')$ is the growth rate $dV/dt$ at time $t''$ of a vapor bubble, created with the critical size at time $t'$. Possible effects of ``cramming'' for many neighboring bubbles, not accounted for in Eq.~(\ref{bc:b_crit0=}), are not expected to be significant because typically $\xi_v \ll 1$.

For large values of the Gibbs number
\begin{equation}\label{bc:G=}
  G =\frac{W_c}{T} \gg 1
\end{equation}
there is no need to calculate the time integral in Eq.~(\ref{bc:b_crit0=}) because any integral of the form $\int_0^t \Phi(t') e^{-G(t')} \, dt'$ can be asymptotically accurately evaluated as
\begin{eqnarray}
  \int\limits_0^t \Phi(t') e^{-G(t')} \, dt'  &\approx &
  \Phi(t) e^{-G(t)} \int\limits_{-\infty}^0 e^{-\tau \dot{G}(t)}d\tau
  \nonumber \\ \label{bc:int=} &=&
  \Phi(t) \left[-\dot{G}(t)\right]^{-1} e^{-G(t)}
\end{eqnarray}
by using the first two terms
\begin{equation}\label{bc:Taylor-G=}
  G(t') \approx G(t) +(t'-t)\dot{G}(t)
\end{equation}
of the Taylor series expansion in the exponent --- provided that the time derivative $\dot{G} \equiv dG/dt <0$. Finally, having applied the approximation (\ref{bc:int=}) to Eq.~(\ref{bc:b_crit0=}), we rewrite our criterion for instantaneous transition from a metastable state $(\rho,T)=(\rho,T_1)$ to the equilibrium state $(\rho,T_2)$ in the form
\begin{equation}
  \begin{aligned}\label{bc:b_crit=}
  NV_c&\left(\frac{3\sigma}{\pi m}\right)^{1/2} \left[ \frac{d}{dt}
  \left(-\frac{W_c}{T} \right)\right]^{-1}
  \exp\left(-\frac{W_c}{T}\right)\\&
  =\frac{\rho_l(T_2)-\rho}{\rho_l(T_2)-\rho_v(T_2)}.
  \end{aligned}
\end{equation}
Note that within the approximation (\ref{bc:int=}) the bubble growth rate $\dot{V}_c(t',t'')$ drops out from the answer which corroborates the premise that the growth of supercritical bubbles can indeed be neglected.

\subsection{Constraints due to applicability limits; the effective  boiling rate}

Formula (\ref{bc:b_crit=}) has been derived under a number of physical approximations, some of which break down near the spinodal and critical point, as well as on very short time scales. As a consequence, to avoid physically absurd results, additional constraints may need to be imposed on Eq.~(\ref{bc:b_crit=}) if it is to be used over the entire metastable region. First of all note that as a $(v,p)$ state approaches the spinodal, the difference $p_e-p$ grows and the critical radius $r_c$ falls. Depending on a particular model of EOS and surface tension $\sigma$ (which normally decreases with the increasing temperature and decreasing bubble radius \cite{Skr74}), one may formally calculate the values of $r_c$ less than the intermolecular distance. In such a case nothing bad happens with the exponential factor in Eq.~(\ref{bc:b_crit=}), which simply approaches unity, but the preexponential factor becomes unphysical: clearly, the number of molecules  inside a critical bubble, given by the product $NV_c$, must at least be larger than one
\begin{equation}\label{bc:NV_c>1=}
  NV_c> 1 ;
\end{equation}
more realistic would probably be limiting values $NV_c\gtrsim 10$--100.

Less clear is a possible limit of $\sigma\to 0$ near the spinodal (whether this should be the general case, still seems to be an open question \cite{Skr74}). It would hardly be reasonable to accept that the nucleation rate $J$, given by Eq.~(\ref{bc:J=}), goes to zero on the spinodal. Then, a satisfactory practical choice might be to use in the preexponential factors in Eqs.~(\ref{bc:J=}) and (\ref{bc:b_crit=}) a fixed (at a given pressure $p$) value of $\sigma =\sigma_m$, measured at the melting point. The latter implies that formally one can reach the spinodal before the $\mbox{MS} \rightarrow \mbox{EQ}$ transition occurs; in practice, however, because of a very short required hydrodynamic time scale, it would mean that the hydrodynamic approach itself is no longer applicable.

Another possibility is that the effective rate of explosive boiling
\begin{equation}
  \begin{aligned}\label{bc:tau_b=}
  \tau_b^{-1}=& \frac{NV_c}{\xi_v}
  \left(\frac{3\sigma}{\pi m}\right)^{1/2}
  \left[ T\frac{\partial}{\partial T}
  \left(-\frac{W_c}{T} \right)\right]^{-1}_p\\&
  \times\exp\left(-\frac{W_c}{T}\right),
  \end{aligned}
\end{equation}
inferred from Eq.~(\ref{bc:b_crit=}), formally goes to infinity in the limit of $\sigma\to 0$ for a fixed value of $NV_c$. In this case one can use the same model for $\sigma$ everywhere because an infinite value of $\tau_b^{-1}$ along the spinodal and at the critical point would be consistent with our principal approximation of instantaneous explosive boiling;  also, it would guarantee that the spinodal itself can never be crossed without a transition to the EQ-EOS.

The effective boiling rate (\ref{bc:tau_b=}) is defined such that for the isobaric heating the criterion (\ref{bc:b_crit=}) is equivalent to
\begin{equation}\label{bc:b_crit2=}
  \tau_h^{-1} \equiv \frac{d\ln T}{dt} = \frac{q}{c_pT} = \tau_b^{-1},
\end{equation}
where $d\ln T/dt$ is the full (Lagrangian) time derivative in a fixed fluid element. In a general case, when the quasi-static approximation is no longer valid, one should use the original criterion (\ref{bc:b_crit=}), where the time derivative in the preexponential factor of  Eq.~(\ref{bc:b_crit=}) is calculated along the actual evolution path of the considered fluid element.

Two different models of surface tension were used in this work to evaluate the effective boiling rate. One is the Guggenheim--Katayama law \cite{Gug45,Kat16}
\begin{equation} \label{bc:sig_1=}
\sigma\left(T\right)=\sigma_m\left(\frac{T_c-T}{T_c-T_m}\right)^{11/9},
\end{equation}
where $T_c$ is the critical temperature, $T_m$ is the melting temperature, and $\sigma_m$ is the surface tension at the melting point. For SiO${}_2$ one has $T_m=2073$~K and $\sigma_m=307$~dyne/cm \cite{Kin59}, while the FEOS model yields $T_c=4868$~K.

As an alternative to the Guggenheim--Katayama formula (\ref{bc:sig_1=}), we also used the dependence
\begin{equation}
  \begin{aligned}
  \sigma\left(T,\rho\right) = & \sigma_m
  \left(\frac{T_c-T}{T_c-T_m}\right)
  \left(\frac{\rho_l(T)- \rho_v(T)}{\rho_l(T_m)-
  \rho_v(T_m)}\right)^{2/3} \\& \label{bc:sig_2=}
  \times\left(\frac{\rho-\rho_{l,sp}(T)}{\rho_l(T)-
  \rho_{l,sp}(T)}\right)^{1/2},
  \end{aligned}
\end{equation}
proposed in Ref.~\cite{PoKh.09} to make the surface tension vanish not only at the critical point but along the entire spinodal; here $\rho_{l,sp}(T)$ is the density along the liquid branch of the spinodal.

\begin{figure}
\centering
\includegraphics*[width=0.9\columnwidth]{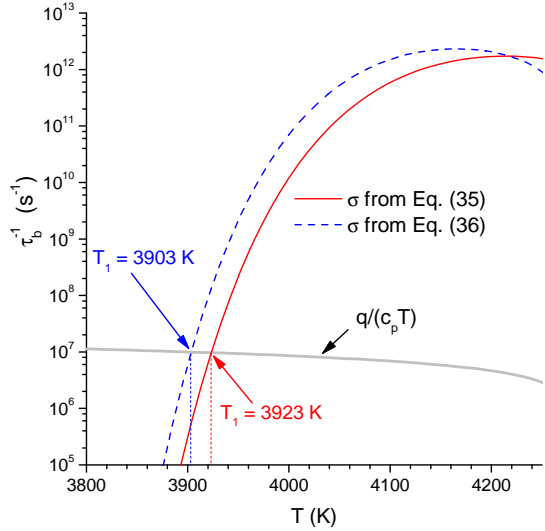}
\caption{\label{f:tau_b} Temperature dependence of the effective rate of explosive boiling $\tau_b^{-1}$, defined in Eq.~(\ref{bc:tau_b=}), for two different formulas for the surface tension $\sigma$ at a fixed pressure $p= 500$~bars.}
\end{figure}

Fig. \ref{f:tau_b} shows the temperature dependence of the effective rate of explosive boiling $\tau_b^{-1}$ of SiO${}_2$, calculated for the two surface tension models (\ref{bc:sig_1=}) and (\ref{bc:sig_2=}); the pressure was fixed at $p= 500$~bars. Both curves rise very steeply over a relatively narrow range of temperatures to the maximum possible values $\tau_{b,max}^{-1} \simeq 10^{12}$~s${}^{-1}$. Intersection with a slowly varying effective heating rate $\tau_h^{-1} = q/(c_pT)$ yields the point of explosive boiling $T=T_1= 3923$~K and 3903~K for, respectively, the Guggenheim--Katayama law and Eq.~(\ref{bc:sig_2=}). This intersection occurs at the value of the Gibbs number $G \simeq 15$, which is large enough to justify the above made approximations. The difference of 20~K between the two $T_1$ values illustrates the accuracy, with which one can calculate this quantity.

In the quasi-static isobaric approximation, all the fluid elements of a uniformly heated planar foil reach the transition temperature $T_1$ and boil up simultaneously --- which significantly simplifies the analysis of the ensuing hydrodynamic picture. In the numerical simulations (which always exhibit small deviations from the quasi-static solution) a simultaneous transition from the MS-EOS to the EQ-EOS was simply enforced over the entire sample once the boiling criterion was fulfilled in its center. The results, discussed below, were obtained with the transition temperature $T_1= 3923$~K, calculated with the Guggenheim--Katayama $\sigma$ law.

\subsection{Subsequent evolution of the two-phase states}

Once a uniform sample undergoes transition to the equilibrium state, all its fluid elements start from the same point $B_2$ on the $p$--$v$ diagram of Fig.~\ref{f:MS->EQ}. But then the trajectories of different fluid elements may diverge because such a transition is typically accompanied by a dramatic (by more than a factor of 10) reduction of the sound speed $c_s$, and the quasi-static approximation may no longer be applicable. As a consequence, the temperature and pressure of fluid elements deep inside the sample would continue to rise because of persistent external heating until the rarefaction wave, propagating from the boundary surface, arrives.

If, however, the condition (\ref{qs:at_s<=}) for quasi-static expansion is still fulfilled, all the fluid elements would quickly (on the time scale $t_s$) relax to the ambient pressure $p=p_0$. If this relaxation occurs with a negligible contribution from the external heating and without shock dissipation, the specific entropy remains constant. Thus, the lower limit to the fluid state trajectories beyond point $B_2$ in Fig.~\ref{f:MS->EQ} corresponds to a quick isentropic expansion to point $B_3$, where the pressure $p_3$ is equal to the initial pressure $p_0$ and the specific entropy $s_3=s_2$ is the same as at point $B_2$, followed by a quasi-static expansion along the equilibrium isobar $p=p_0$. This limiting trajectory would also correspond to the fluid elements at the outer boundary kept at the fixed pressure $p_0$.

\section{\label{s:res} Hydrodynamics of a planar SiO${}_2$ foil undergoing explosive boiling}

\subsection{Numerical setup}

One-dimensional hydrodynamic simulations were performed  with the CAVEAT code \cite{CAVEAT} for a planar layer of SiO${}_2$ with the initial thickness of $l_0=10$~$\mu$m. The initial thermodynamic state was taken at $T_0= 312$~K, $p_0= 500$~bar, which corresponds to the initial density of $\rho_0 = 2.20$~g~cm${}^{-3}$ and the sound speed $c_s =4.56\times 10^5$~cm/s. Starting from $t=0$, a constant volumetric heating at a rate $q= 10^{11}$~J~g${}^{-1}$~s${}^{-1}$ was applied for the period of $0< t< 150$~ns. Since the initial sonic time $t_s=l_0/c_s \approx 2.2$~ns is short compared to the heating time, the preconditions for quasi-static heating are fulfilled. To suppress the unwanted quasi-elastic oscillations, the initial velocity profile was set according to the quasi-static solution (\ref{qs:u_q=}). The foil was simulated under a fixed boundary pressure equal to its initial value $p_0= 500$~bar, which enabled a much clearer visual representation of the boundary fluid-element trajectories in Figs.~\ref{f:vp-tracks_MS} and \ref{f:vp-tracks_EQ} below.

The SiO${}_2$ EOS was calculated by using the FEOS package (\ref{s:MPQeos}), where the solid part of the ionic contribution was omitted for simplicity. Also, the following input parameters were used: $\rho_o=2.2$~g~cm${}^{-3}$, $T_o=300$~K, and $K_o=37$~GPa (the bulk modulus). The parameters of the soft-sphere model, $\epsilon_{coh}= 9.74$~kJ/g, $m=0.8$, and $n=2.8$,  were adjusted such as to make the critical point approach the data by Iosilevskiy and Gryaznov \cite{IoGr05,IoHy.01}: $T_c= 6300$~K, $p_c= 3.5$~kbar, and $\rho_c = 0.65$~g~cm${}^{-3}$.

\subsection{Simulation results}

\begin{figure}[t]
\centering
\includegraphics*[width=\columnwidth]{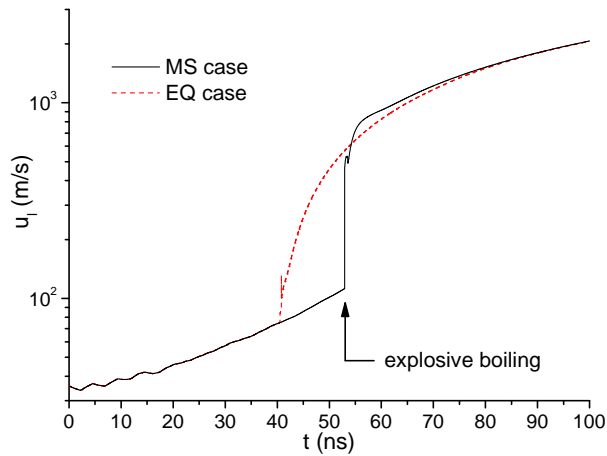}
\caption{\label{f:u_l(t)} Temporal profile of the surface velocity $u_l(t)$ of the uniformly heated SiO$_2$ foil in two cases: obeying the metastable EOS until the explosive boiling occurs at $t_{eb}= 52.95$~ns (MS case), and obeying the equilibrium EOS immediately after crossing the binodal (EQ case).}
\end{figure}

Two cases were simulated: (i)~the MS case, where SiO${}_2$ obeys the metastable EOS until the temperature $T_1= 3923$~K of explosive boiling is reached at $t= t_{eb} =52.95$~ns, after which a jump to the equilibrium EOS has been made, and (ii)~the EQ case, where the continuous transition to the EQ-EOS is made immediately after crossing the binodal. Fig. \ref{f:u_l(t)} shows the evolution of the surface velocity $u_l(t)$ for both cases. The initial oscillations could not be fully suppressed by assigning the quasi-static initial velocity profile (\ref{qs:u_q=}) because the parameter $at_s$ had a finite value.

As might be expected, in the MS case one observes a strong jump in the surface velocity $u_l$ at the moment of explosive boiling due to a sudden increase of pressure from $p_0= 0.5$~kbar to $p_2= 1.6$~kbar (the temperature jumps by 40~K to $T_2= 3964$~K). In the EQ case, on the contrary, the surface velocity rises rather gradually on a time scale of $\Delta t \simeq 15$~ns. Note that, once the artificial boundary pressure of $p_0 =500$~bar is lowered, the velocity rise in the EQ case becomes even more gradual than shown in Fig.~\ref{f:u_l(t)}. We expect that such a significant difference in temporal behavior of the surface velocity of volumetrically heated samples should be definitely differentiable in corresponding experiments; also, it clearly demonstrates the necessity of adequate modeling of the explosive boiling for interpretation of such experiments.

\begin{figure}[t]
\centering
\includegraphics*[width=\columnwidth]{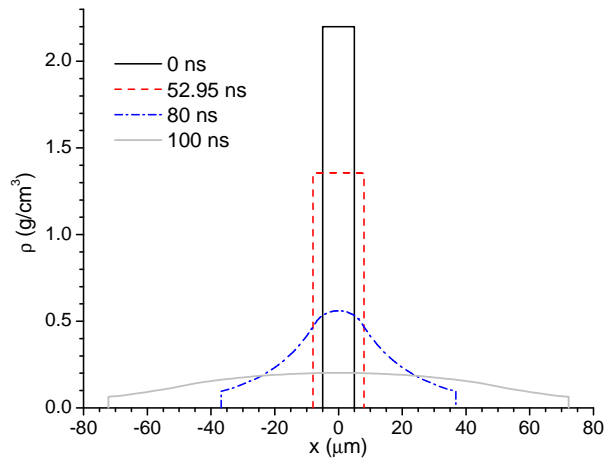}
\caption{\label{f:rho} Density profiles in the SiO$_2$ foil at four characteristic times in the MS case.}
\end{figure}

Fig. \ref{f:rho} presents the evolution of the density profile across the foil in the MS case. Prior to the moment of explosive boiling $t= 52.95$~ns the liquid foil is bounded by sharp surfaces and has a box-like density profile. Having boiled up, it expands as a planar layer of gas: had the boundary pressure been set to zero, the spatial density profile would have approached the Gaussian shape, characteristic of the self-similar regime with a uniform temperature and linear velocity distributions \cite{LoRo86}.

\begin{figure}[t]
\centering
\includegraphics*[width=0.9\columnwidth]{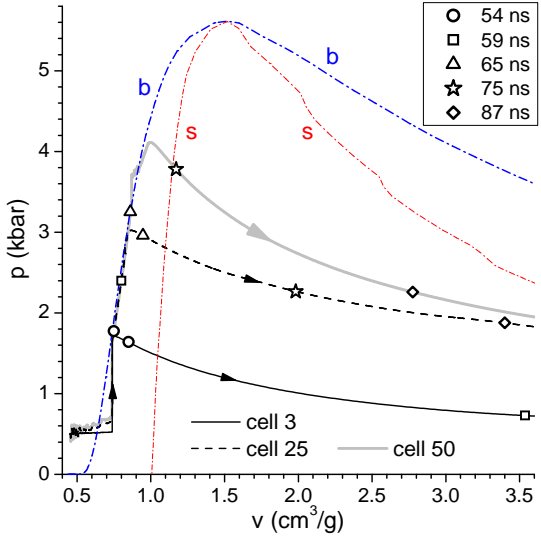}
\caption{\label{f:vp-tracks_MS} Trajectories of three selected Lagrangian cells of the simulated SiO$_2$ foil on the $(v,p)$  phase plane in the MS case. Evolution in time is indicated with a series of time markers.}
\end{figure}

\begin{figure}[t]
\centering
\includegraphics*[width=0.9\columnwidth]{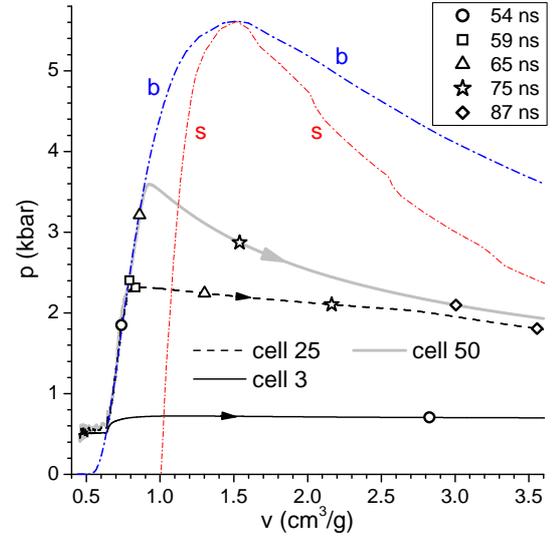}
\caption{\label{f:vp-tracks_EQ} Same as Fig.~\ref{f:vp-tracks_MS} but for the EQ case.}
\end{figure}

Figs. \ref{f:vp-tracks_MS} and \ref{f:vp-tracks_EQ} display the trajectories of three selected Lagrangian cells on the thermodynamic $(v,p)$ plane as calculated, respectively, in the MS and EQ cases. Initially, the foil thickness was divided into 100 equal mesh intervals, and the simulations were performed in the purely Lagrangian mode. Hence, cell 50 corresponds to the mid-plane of the foil, cell 25 -- to the half-distance (along the mass coordinate) between the mid-plane and the edge, and cell~3 lies near the foil edge.

Qualitatively, the $v$--$p$ trajectories in Fig.~\ref{f:vp-tracks_MS} are similar to those depicted schematically in Fig.~\ref{f:MS->EQ}. Up to the very moment of explosive boiling, all the fluid elements follow roughly the same quasi-isobaric path, but begin to diverge sharply after the jump to the equilibrium state with $p= p_2 =1.65$~kbar. While the liquid--vapor mixture near the foil edge rapidly expands, with pressure relaxing to the boundary value of $p_0=0.5$~kbar, the pressure in deep interior of the foil continues to rise up to $p_{max} \approx 4$~kbar until the rarefaction wave reaches the foil center.

A salient feature in Figs.~\ref{f:vp-tracks_MS} and \ref{f:vp-tracks_EQ} is that, after crossing the binodal but before arrival of the rarefaction wave, the $v$--$p$ trajectories of the inner foil elements virtually stick to the binodal. Such behavior is caused by a large drop in the value of the sound velocity across the binodal, namely, by a factor ranging from 30 to 3 in the relevant pressure interval. Because the sonic time inside the binodal becomes large (tens of nanoseconds), one would expect that the pressure in the heated foil center should rise practically vertically on the $v$--$p$ diagram (i.e.\ at a practically constant density). But then the corresponding $v$--$p$ trajectory would return into the pure liquid phase outside the binodal --- where, however, the sonic time is short (a few nanoseconds) and the foil material would be forced to rapidly reenter the two-phase region along a quasi-isobaric path. As a result, for about 20~ns the binodal becomes sort of an attractor for the $v$--$p$ trajectories of the fluid elements near the foil center.

Finally note that at $t \gtrsim 100$~ns the $v$--$p$  trajectories from the two simulations become quite close to each other, i.e.\  the final state of the evaporated foil does not depend strongly on whether one uses the metastable or the equilibrium EOS in the two-phase region. It is the dynamical behavior inside the two-phase region which requires adequate modeling of the boiling (condensation) process in the superheated liquid (supercooled vapor).

\section{Conclusion \label{s:c}}

In this work we investigate a  physical model aimed at a self-consistent description of the dynamics of matter evolving through the two-phase region of the phase diagram, where the liquid and the vapor states can coexist. In particular, we analyze in detail the case of volumetric heating of a liquid that should eventually result in its full vaporization. The proposed model is based on the assumption that in the two-phase region the heated liquid should be described with the metastable EOS until the criterion of explosive boiling is fulfilled, at which point an instantaneous irreversible transition to the equilibrium EOS (obtained by applying the Maxwell rule) can be made. Such an approach appears to be the simplest possible and still physically self-consistent in the framework of pure hydrodynamics, where no additional kinetic equations for description of phase transitions are to be solved.

Starting from the theory of homogeneous bubble nucleation in superheated liquids, we formulate the criterion for explosive boiling (\ref{bc:b_crit=}) that has the advantage of being local in space and time. We argue that the irreversible jump to the equilibrium EOS must be made under the conditions of constant density and specific internal energy, and show that such a jump is accompanied by a step-like increase of pressure and temperature. The only quantity beside EOS that enters the boiling criterion (\ref{bc:b_crit=}) is the surface tension of liquid $\sigma$, whose dependence on temperature and density is assumed to be known.

The main properties of the proposed model are illustrated on a simple example of a uniformly heated planar layer of SiO${}_2$. Hydrodynamic simulations of this layer have demonstrated that adequate modeling of the transition from the metastable to the equilibrium state can be rather important for correct interpretation of experimental data when, for example, the thermal expansion of such foils is measured.

In the present work the proposed method of hydrodynamic treatment of the two-phase region was applied to a perfectly uniform sample, where the boiling transition occurs simultaneously over the entire volume. However, the local transition criterion (\ref{bc:b_crit=}) can of course be used in situations where the distribution of thermodynamic parameters is non-uniform (or slightly non-uniform). In such cases one may observe a qualitatively different behavior: small initial non-uniformities can be strongly enhanced by sudden local jumps in pressure. In reality this phenomenon may manifest itself as a spontaneous generation of foam-like structures --- like it has been observed by a quasi-uniform Joule heating of exploding wires \cite{PiSh.99}. But a detailed analysis of such behavior remains for future work.

As a word of caution, it should be noted that, having based our main criterion (\ref{bc:b_crit=}) on the theory of homogeneous nucleation, we have tacitly assumed that possible influence of impurities and/or boundary conditions can be ignored. Very often this may not be the case, and then a more complex analysis, based for example on the theory of heterogeneous nucleation may be needed.

\section*{Acknowledgments \label{s:acknow}}
This work was supported by the Extreme Matter Institute EMMI, the Bundesministerium f\"ur Bildung und Forschung BMBF (Projects 06FY9085 and 06FY9080I), the Joint Institute for High Temperature of RAS Scientific Program ``Physics of extreme states of matter'', and by the Moscow Institute of Physics and Technology Education Center ``Physics of High Energy Density Matter''.

\appendix

\section{\label{s:MPQeos} EOS model FEOS}

The ``Frankfurt equation-of-state package (FEOS)'' is a C++ computer code that can generate EOS tables for arbitrary materials. It is an improved version of the code MPQeos, which is described in the corresponding MPQ Report \cite{KeMtV98}. The underlying physical model is the ``quotidian equation-of-state model (QEOS)'' \cite{MoWa.88}. A more detailed documentation on the FEOS package can be found at http://th.physik.uni-frankfurt.de/$\sim$faik/.

\subsection*{The MPQeos model}

The original MPQeos code can only generate EOS of chemical elements. Thermodynamic quantities, like pressure, specific internal energy, specific entropy, are derived from the specific Helmholtz free energy $f=\epsilon-Ts$, which is composed of three contributions: the electronic part, the ionic part, and the semiempirical bonding correction. All three components of $f$ are assumed to be a function of a single temperature $T$, although in principle there is no difficulty in separating the ion and the electron temperatures. For any particular element, the MPQeos model requires two empirical parameters to calibrate the generated EOS, namely, the density $\rho_o$ and the bulk modulus $K_o=\rho\left(\partial p/\partial\rho\right)$ at a certain reference value of temperature $T_o$ and zero pressure $p=p_o=0$. For substances that are in the solid state at normal conditions, usually the values $T_o \approx 300$~K are used.

\paragraph*{Electronic contribution}
The electronic contribution is calculated by using the Thomas--Fermi (TF) model \cite{Tho27,Fer28}. Matter is envisaged as consisting of elementary spherical cells, for which the equilibrium electronic distribution --- and therewith also the thermodynamic quantities of the electrons --- is calculated by solving the TF equation. An important advantage of the TF model is the fact that the TF equation and all thermodynamic quantities scale with the atomic number $Z$, and therefore must be calculated only once, e.g.\ for hydrogen. The most important drawback of the simple TF model is the absence of attractive (bonding) forces between neutral atoms. Hence, pressures near the normal conditions are overestimated.

\paragraph*{Bonding correction}
MPQeos uses a bonding correction from Ref.~\cite{Bar67}. It adjusts the EOS at a reference point $(p,T)=(0,T_o)$ to have the correct (or user-defined) values of the density $\rho_o$ and the bulk modulus \cite{ZeRa02} $K_o=\rho\left(\partial p/\partial\rho\right)$. The reference temperature $T_o$ is usually chosen such that at $p=0$ the studied material is in the solid state --- i.e.\ well below the critical temperature $T_c$. A relatively small difference between the isothermal and isentropic bulk moduli is usually ignored. The bonding correction has the strongest effect on EOS  near the solid density. At higher densities the TF electronic contribution dominates.

\paragraph*{Ionic contribution}
Contribution of ions to thermodynamic quantities is fully independent of that from electrons. It is calculated by using the Cowan model as described in Ref.~\cite{MoWa.88}, which employs analytical formulas to smoothly interpolate between the Debye solid, the normal solid and the liquid states. The ionic internal energy and pressure are of purely thermal nature.

\subsection*{Modifications in the FEOS model}

Three important changes were introduced into the original MPQeos code, which have lead to the improved version FEOS.

\paragraph*{Cold curve improvement}
The QEOS model overestimates the critical pressure $p_c$ and the critical temperature $T_c$. This problem was solved by replacing the TF cold curve and the bonding correction for densities $\rho< \rho_o$ with a soft-sphere function proposed by Young and Corey \cite{YoCo95}
\begin{equation}
  \epsilon_{cold}\left(\rho,T=0\right)=A\rho^n-B\rho^m+\epsilon_{coh}.
\end{equation}
Here $\epsilon_{coh}$ is the cohesive energy, or the enthalpy of sublimation per unit mass. The constants $A$ and $B$ are adjusted so as to make the total pressure and the internal energy be equal to zero at the reference point $(\rho_o,T_o)$:
\begin{equation}\label{EOS:p_o=0}
  p(\rho_o,T_o)=\epsilon(\rho_o,T_o)=0.
\end{equation}
The free parameters $m$ and $n$ are used to improve the agreement with the experimentally (or theoretically) known critical point. At $\rho = \rho_o$ the sound velocity is allowed to be discontinuous.

\paragraph*{Calculation of mixtures of elements}
A well known rule for the TF mixing of elements \cite{MoWa.88} was added to the FEOS. The ionic contribution and the bonding correction are handled as a single species with a mean atomic number $\bar{Z}$ and weight $\bar{A}$. For the electronic contribution of a mixture the partial volumes of all species $k$ are iteratively adjusted in order to equilibrate the TF pressures $p_{e,k}$ and to fulfill an additive volume rule. Finally, the appropriate combination of the electronic free energies is calculated.

\paragraph*{Calculation of liquid--vapor phase coexistence}
In the original MPQeos the liquid--vapor phase coexistence  was calculated for each isotherm below the critical temperature $T_c$ by using Maxwell's geometrical rule \cite{LL-sp96}. Especially for low temperatures $T\ll T_c$, this rule is computationally very intensive and imprecise. In the FEOS this problem was solved by applying a new routine, where the saturated vapor pressure is calculated by equating the values of the Gibbs' free energy and pressure on the liquid and vapor branches of the binodal for each isotherm below $T_c$.

\section{\label{s:Ac} Acoustic oscillations of a planar foil}

Consider slow subsonic ($u\ll c_s$) one-dimensional motion in a planar foil, which has no heat exchange with external sources ($q=0$) and whose equilibrium state at a fixed boundary pressure $p_0$ corresponds to a uniform density $\rho_0$; the equilibrium foil thickness is $l_0$. Then the density and the pressure perturbations are relatively small and can be cast in the form
\begin{equation}\label{Ac:rho,p=} \begin{aligned} &
  \rho(t,x) \approx\rho_0+\rho_1(t,x) = \rho_0 [1+\eta(t,x)],\\ &
  p(t,x) \approx p_0 +\left(\frac{\partial p}{\partial\rho}\right)_s
  \rho_1= p_0+\rho_0 c_s^2 \eta(t,x), \end{aligned}
\end{equation}
where $|\eta| \ll 1$. Substituting (\ref{Ac:rho,p=}) into the hydrodynamic Eqs. (\ref{qs:hyd1=}) and  (\ref{qs:hyd2=}) and retaining only the first-order terms with respect to the small quantities $\eta$ and $u$, we obtain
\begin{equation}\label{Ac:du/dt=} \begin{aligned} &
  \frac{\partial \eta}{\partial t} +
  \frac{\partial u}{\partial x} =0, \\ &
  \frac{\partial u}{\partial t} + c_s^2
  \frac{\partial \eta}{\partial x} =0, \end{aligned}
\end{equation}
which yield the classical wave equation
\begin{equation}\label{Ac:wave_eq=}
  \frac{\partial^2 \eta}{\partial t^2} =c_s^2
  \frac{\partial^2 \eta}{\partial x^2}, \quad
  \frac{\partial^2 u}{\partial t^2} =c_s^2
  \frac{\partial^2 u}{\partial x^2}
\end{equation}
for both $\eta$ and $u$.

Because the pressure is fixed at the left and right boundaries $x = \pm l_0/2$, i.e.\ $\eta(t,\pm l_0/2) =0$, our problem is mathematically fully equivalent to the classical problem of string vibrations \cite[Ch.~V, \S~3]{CouHi89}. The general solution is given by a superposition
\begin{equation}\label{Ac:eta_o=}
  \eta \equiv \eta_o(t,x) = -\sum\limits_{n=1}^{\infty}
  \frac{u_{on}}{c_s}
  \sin(\omega_n t +\phi_n) \, \cos\left(\frac{\omega_n x}{c_s}\right),
\end{equation}
\begin{equation}\label{Ac:u_o=}
  u \equiv u_o(t,x) = \sum\limits_{n=1}^{\infty} u_{on}
  \cos(\omega_n t +\phi_n) \, \sin\left(\frac{\omega_n x}{c_s}\right),
\end{equation}
of standing acoustic waves with a discrete set of eigenfrequencies
\begin{equation}\label{Ac:omeg_n=}
  \omega_n= \frac{\pi c_s}{l_0} (2n-1), \quad n=1,2,\ldots ,
\end{equation}
where $\phi_n$ and $u_{on}$ are arbitrary constants. The oscillatory solution (\ref{Ac:eta_o=}), (\ref{Ac:u_o=}) is periodic in time, with the period
\begin{equation}\label{Ac:t_o}
  t_o = \frac{2\pi}{\omega_1} = \frac{2l_0}{c_s} =2t_s,
\end{equation}
given by the fundamental frequency $\omega_1$.

Now, if we consider the case of a sudden turnon of the heating rate $q=q_m$ at $t=0$ under the condition (\ref{qs:at_s<=}), the general first-order solution for the fluid velocity $u$ can be obtained as a sum of the quasi-static and oscillation terms,
\begin{equation}\label{Ac:u=}
  u(t,x) = ax +u_o(t,x).
\end{equation}
For $at_s <<1$ simple addition of the two linear solutions is justified so far as the relative changes of $a=\Gamma q_m/c_s^2$ and $c_s$ remain small, i.e.\ at least over a certain time interval $0< t \ll a^{-1}$, which may still be large compared with the oscillation period $t_o$.

With the initial condition $u(0,x)=0$ we must have $u_o(0,\pm l_0/2)=\mp \frac{1}{2}al_0$ and the coefficients $u_{on}$ (at least some of them) in Eq.~(\ref{Ac:u_o=}) will be of the order of $|u_{on}| \simeq al_0$. By virtue of Eq.~(\ref{Ac:eta_o=}), this implies that the spatial variation of the density
\begin{equation}\label{Ac:del_rho=}
  \delta\rho \simeq \rho_0|\eta| \simeq \rho_0\frac{al_0}{c_s} =
  \rho_0\, (at_s)
\end{equation}
is one order lower in the small parameter $at_s$ --- i.e.\ a factor $(at_s)^{-1} \gg 1$ higher --- than the spatial density variation (\ref{qs:del_rho=}) in the quasi-static solution. Hence, to obtain a self-consistent first-order solution, we must assign a zero initial value $\eta_o(0,x)=0$ for the oscillatory component of the density variation. From this we obtain $\phi_n=0$ for all $n=1,2,\ldots$, whereas the values of $u_{on}$ are uniquely determined by the equation
\begin{equation}\label{Ac:u_on=}
  -ax =\sum\limits_{n=1}^{\infty} u_{on}
  \sin\left(\frac{\omega_n x}{c_s}\right).
\end{equation}
Finally, because at $t=0$ we have $d u_o(0,l_0/2)/dt =0$, the amplitude of the oscillatory term in (\ref{Ac:u=}) is
\begin{equation}\label{Ac:u_osc=}
  u_{osc} =\left|u_o(0,l_0/2)\right| =\frac{1}{2} al_0.
\end{equation}

\section{Temperature and entropy jumps by explosive boiling of a superheated liquid \label{s:T2}}

To determine the direction of temperature change by quasi-instantaneous explosive boiling, consider the metastable, $\epsilon_{MS}(v,T)$, and the equilibrium,  $\epsilon_{EQ}(v,T)$, specific internal energies as functions of the specific volume $v$ and the temperature $T$. Clearly, if the heat capacity $c_{v,EQ} = \partial\epsilon_{EQ}/\partial T$ is positive and the inequality
\begin{equation}\label{T2:eps_EQ<eps_MS=}
 \epsilon_{MS}(v,T) >  \epsilon_{EQ}(v,T)
\end{equation}
holds over the entire two-phase region [i.e.\ within the interval $v_l(T)< v< v_v(T)$], then the equation
\begin{equation}\label{T2:eps_EQ=eps_MS=}
  \epsilon_{MS}(v,T) =\epsilon_{EQ}(v,T_2)
\end{equation}
implies $T_2> T$.

\begin{figure}
\centering
\includegraphics*[width=\columnwidth]{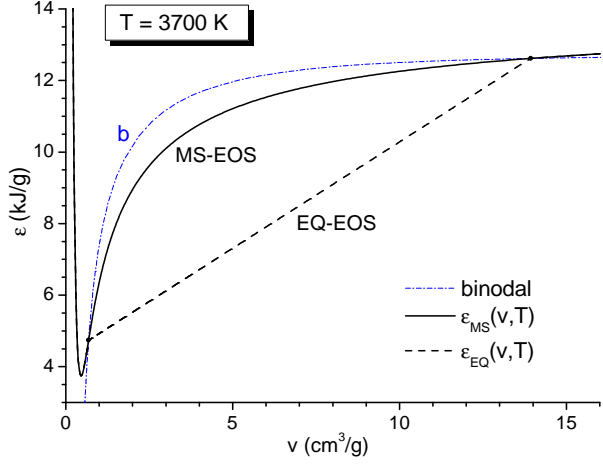}
\caption{\label{f:eps_MS_EQ} Metastable (solid) and equilibrium (dashes) isotherms for $T=3700$~K on the $(v,\epsilon)$ plane for SiO${}_2$ in the two-phase region as calculated with the FEOS model. Validity of inequality (\ref{T2:eps_EQ<eps_MS=}) is clearly observed.}
\end{figure}

To prove (\ref{T2:eps_EQ<eps_MS=}), consider the function
\begin{equation}\label{T2:phi=}
  \phi(v)=\epsilon_{MS}(v,T)-\epsilon_{EQ}(v,T)
\end{equation}
for some fixed value of $T$. This function vanishes by definition on the binodal, i.e.\ $\phi(v_l(T)) =\phi(v_v(T))=0$. If, in addition, we have
\begin{equation}\label{T2:phi''<0=}
  \frac{d^2\phi}{dv^2} < 0 \quad \mbox{for all} \quad v_l(T)< v< v_v(T),
\end{equation}
we necessarily obtain (\ref{T2:eps_EQ<eps_MS=}). Since $\epsilon_{EQ}(v,T)$ is a linear function of $v$ for any fixed $T<T_c$, we have $\partial^2 \epsilon_{EQ}(v,T)/ \partial v^2 =0$, and an inequality
\begin{equation}\label{T2:d2eps_EQ=0=}
  \frac{\partial^2 \epsilon_{MS}(v,T)}{\partial v^2}
  \equiv \frac{\partial}{\partial v} \left[
  T\left(\frac{\partial p}{\partial T}\right)_v -p\right] <0
\end{equation}
will necessarily imply $T_2>T$ in Eq.~(\ref{T2:eps_EQ=eps_MS=}). For the van der Waals EOS one readily ascertains (\ref{T2:d2eps_EQ=0=}) by calculating
\begin{equation}\label{T2:vdW_d2e/dv2=}
  \frac{\partial^2 \epsilon(v,T)}{\partial v^2} = -\frac{6p_cv_c^2}{v^3} <0,
\end{equation}
where index $c$ marks the values at the critical point. The FEOS used in this work also has this property, as is illustrated in Fig.~\ref{f:eps_MS_EQ}.

Similarly, if one chooses the specific volume $v$ and the specific entropy $s$ as the independent variables and ascertains that
\begin{equation}\label{T2:eps_EQ(s)<eps_MS(s)=}
 \epsilon_{MS}(v,s) > \epsilon_{EQ}(v,s)
\end{equation}
over the two-phase region, then the equality
\begin{equation}\label{T2:eps_EQ(s)=eps_MS(s)=}
  \epsilon_{MS}(v,s) =\epsilon_{EQ}(v,s_2)
\end{equation}
would automatically imply $s_2> s$ because $\left( \partial \epsilon/\partial s \right)_v = T >0$. Again, (\ref{T2:eps_EQ(s)<eps_MS(s)=}) can be proven for the van der Waals EOS by considering the function
\begin{equation}\label{T2:phi_s=}
  \phi(v)=\epsilon_{MS}(v,s)-\epsilon_{EQ}(v,s)
\end{equation}
at some fixed value of $s$. Both $\phi(v)$ and its first derivative
\begin{equation}\label{T2:phi'_s=}
    \phi'(v) \equiv \frac{d\phi}{dv} =p_{EQ}(v,s)-p_{MS}(v,s)
\end{equation}
by definition vanish on the binodal, $\phi(v_l(s)) =\phi'(v_l(s))=0$, $\phi(v_v(s)) =\phi'(v_v(s))=0$. Having recalled the general thermodynamic identity \cite{LL-sp96}
\begin{equation}\label{T2:dp/dv_s=}
    \left(\frac{\partial p}{\partial v}\right)_s =\left(\frac{\partial p}{\partial v}\right)_T - \frac{T}{c_v}\left(\frac{\partial p}{\partial T}\right)_v^2
\end{equation}
and taking into account that for the van der Waals EOS $c_{v,EQ} =c_{v,MS}$ (because $c_v$ is a function of temperature only), one can prove that
\begin{equation}\label{T2:phi''_s=}
   \frac{d^2\phi}{dv^2} >0
\end{equation}
in the metastable region from the liquid side, i.e.\ at $v>v_l(s)$ but before the spinodal. Then, since  $\phi'(v)$ can cross zero not more than once within the interval $v_l(s)< v<v_v(s)$, we necessarily have $\phi(v)>0$ everywhere in the two-phase coexistence region. The fact that for normal fluids (as contrasted to the so called retrograde fluids) $v_l(s)= \infty$ (i.e.\ that normally both the MS and the EQ isentropes on the $(v,p)$ plane never actually exit the two-phase coexistence region) does not affect the above argumentation for the particular case of the van der Waals EOS.

%\section*{References}
\bibliographystyle{model1-num-names}
\bibliography{LVhyd2012_final}

\end{document}